\documentclass{elsart}

\def\ggs{\buildrel\textstyle > \over {\hbox{\raise0.05ex\hbox{$\sim$}}}}
\def\lls{\buildrel\textstyle < \over {\hbox{\raise0.05ex\hbox{$\sim$}}}}
\def\gsim{\,\lower0.95ex\hbox{$\ggs$}\,}
\def\lsim{\,\lower0.95ex\hbox{$\lls$}\,}

\usepackage{graphicx}
\usepackage{amssymb}

\begin{document}

\begin{frontmatter}

\title{Ising transition in a one-dimensional
   quarter-filled electron system with dimerization}

 \author[label1]{M.~Tsuchiizu}\ead{tsuchiiz@yukawa.kyoto-u.ac.jp},
 \author[label2]{E.~Orignac}
 \address[label1]{Yukawa Institute for Theoretical Physics, 
                 Kyoto University,  Kyoto 606-8502, Japan}
 \address[label2]{Laboratoire de Physique Th\'eorique, CNRS UMR 8549 
     Ecole Normale Sup\'erieure, 24 Rue Lhomond 75005 Paris, France}

\begin{abstract}
We examine critical properties of the 
   quarter-filled one-dimensional Hubbard model with dimerization and
   with the onsite and nearest-neighbor Coulomb repulsion $U$ and $V$.
By utilizing the bosonization method, 
   it is shown that the system exhibits an Ising quantum phase 
   transition from the Mott insulating state 
   to the charge-ordered insulating state.
It is also shown that 
   the dielectric permittivity exhibits a strong enhancement 
   as decreasing temperature with power-law dependence
   at the Ising critical point.
\end{abstract}

\end{frontmatter}


The Bechgaard salts (TMTSF)$_2$X and the Fabre salts 
   (TMTTF)$_2$X form a family of organic conductors with 
  quasi-one-dimensional band structure, 
  with a long history of both theoretical 
  and experimental research. 
These materials have a 3/4-filled electron (i.e., quarter-filled hole) 
   band due to  
   a charge transfer from two organic
   molecules to one counterion X.
Since  the
    transfer integral is slightly dimerized along the chain axis,
   the system becomes effectively half-filling.  
Optical and photoemission experiments on the normal states 
   have shown evidences of a
   charge gap for both TMTTF salts and TMTSF salts
   \cite{Gruner_ICSM,Gruner_science}. This charge gap is believed to
be related to the formation of a uniform insulating state. 
Recently,  a charge ordered (CO) insulating state
   have been discovered by  NMR measurements
   in the (TMTTF)$_2$AsF$_6$ and (TMTTF)$_2$PF$_6$ salts \cite{Chow} 
   and by  dielectric permittivity measurements
   in the (TMTTF)$_2$PF$_6$ salt \cite{Monceau}.

In order to describe theoretically the insulator,
   one has to take into account the commensurability effect
   of both half-filling and quarter-filling
   \cite{Giamarchi_physica}.
As a simplified model,
   the quarter-filled one-dimensional (1D)
   extended Hubbard model with dimerization and 
   with the onsite and nearest-neighbor Coulomb repulsion
   has been proposed
   and it is shown in a mean-field approach that
   a transition from the 
   state with a uniform charge distribution
   to the CO state takes place
   by increasing the intersite Coulomb repulsion \cite{Seo}.
In order to obtain the characteristics of such a transition,
   the effect of quantum fluctuations must be taken into account.
For this purpose, the bosonization method is applied to this
   system \cite{Bourbonnais,Tsuchiiz}.
The model reduces to the 
   double-frequency sine-Gordon (DSG) model 
   for the charge degrees of freedom    
   where two kinds of non-linear terms
   appear due to the half-filled and quarter-filled 
   umklapp scattering.
The DSG model has been analyzed classically \cite{classical_DSG},
   and quantum field theoretically 
   \cite{Delfino,Fabrizio_PRL,Fabrizio,Tsuchiiz}. 
In terms of the renormalization group (RG) method,
   it is shown that there are two-kinds of different 
   fixed points corresponding to the
   Mott insulator and the CO insulator
   \cite{Tsuchiiz}.
However, in the perturbative RG scheme
   the both fixed points are strong coupling
   where the perturbative RG treatment breaks down, thus 
   the low energy states behavior and
   the critical behavior of the transition 
   cannot be captured correctly.

Recently, it has been established
   that
   the critical properties of the DSG model is described as a quantum
   Ising transition, i.e.,
   $c=1/2$ CFT \cite{Delfino,Fabrizio_PRL,Fabrizio}.
By expanding the previous analysis of the DSG model, we examine
   the critical properties of the 
   quarter-filled 1D extended Hubbard model with dimerization,
   and also examine
   the physical quantities, such as dielectric response and 
   optical conductivity, in the present study. Note that in
\cite{Fabrizio_PRL,Fabrizio}, the related problem of the conductivity
in a 1D Hubbard model with a {\it staggered} potential has been
considered. 
We also discuss the experimental results on the
   huge enhancement of the dielectric permittivity
   obtained in the
   (TMTTF)$_2$X salts \cite{Monceau}.

We consider a 1D extended Hubbard model with 
   quarter-filling, where the corresponding Hamiltonian is given by
\begin{eqnarray}
H &=& - \sum_{j,\sigma,l}
   \left[ t + (-1)^{j} t_d\right] 
   \left(  c_{j, \sigma}^{\dagger} 
    \, c_{j+1,\sigma} + \mbox{\rm h.c.} \right)
\nonumber \\ &&{}
+ U \sum_{j} \, n_{j, \uparrow} \, n_{j, \downarrow} 
+ V \sum_{j} \, n_{j} \, n_{j+1} 
\label{eq:H}
\end{eqnarray}
 where 
   $n_{j, \sigma} = c_{j, \sigma}^\dagger c_{j, \sigma}$ and
   $c_{j, \sigma}$ denotes the annihilation operator of the electron
   at the $j$-th site with spin 
   $\sigma$($=\uparrow,\downarrow$), and 
   $t_d(>0)$ 
   denotes dimerization along the chain.
The onsite and nearest-neighbor Coulomb repulsions
   are denoted by $U$ and $V$, respectively.
By utilizing the bosonization method,
   the phase Hamiltonian is given by $H=H_c+H_s$ with
   \cite{Tsuchiiz}
\begin{eqnarray}
H_c &=& 
\frac{v_\rho}{4\pi} \int {\rm d}x 
  \left[\frac{1}{K_\rho} \left(\partial_x \theta_+ \right)^2
         + K_\rho \left(\partial_x \theta_- \right)^2
  \right]
\nonumber \\
&& {}
- \frac{g_{1/2}}{2 \pi^2 \alpha^2} \int {\rm d}x \, \sin 2 \theta_+
+ \frac{g_{1/4}}{2 \pi^2 \alpha^2} \int {\rm d}x \, \cos 4 \theta_+
,
\label{eq:phaseHamiltonian}
\\
H_s &=& 
\frac{v_\sigma}{4\pi} \int {\rm d}x 
  \left[\frac{1}{K_\sigma} \left(\partial_x \phi_+ \right)^2
         + K_\sigma \left(\partial_x \phi_- \right)^2
  \right]
\nonumber \\
&& {}
+ \frac{g_{s}}{2 \pi^2 \alpha^2} \int {\rm d}x \, \cos 2 \phi_+
,
\label{eq:phaseHamiltonian_spin}
\end{eqnarray}
   where $\theta_\pm$ and $\phi_\pm$ are charge and spin 
   phase variables \cite{Tsuchiiz} and
   $\alpha$ is a cutoff 
   of the order of the lattice constant.
In Eq.~(\ref{eq:phaseHamiltonian}),
   the velocity and the Tomonaga-Luttinger exponent are
   given by
   $v_\rho = v_{\rm F} 
    [(1+g_{4\rho}/2\pi v_{\rm F})^2-(g_\rho/2\pi v_{\rm F})^2]^{1/2}$,
    $v_\sigma = v_{\rm F} 
    [(1+g_{4\sigma}/2\pi v_{\rm F})^2
         -(g_\sigma/2\pi v_{\rm F})^2]^{1/2}$,
    $K_\rho = 
    [(2\pi v_{\rm F}+g_{4\rho}-g_\rho)/
           (2\pi v_{\rm F}+g_{4\rho}+g_\rho)]^{1/2}$
    and
    $K_\sigma = 
    [(2\pi v_{\rm F}+g_{4\sigma}-g_\sigma)
     /(2\pi v_{\rm F}+g_{4\sigma}+g_\sigma)]^{1/2}$.
 The constants $g_\rho$, $g_{4\rho}$, $g_\sigma$ and $g_{4\sigma}$
    are given by 
   $g_{\rho} =g_{4\rho} = (U + 4V)a$ and
   $g_{\sigma} =g_{4\sigma} = Ua$
   in the lowest order of the interaction, where
   $a$ is the lattice constant.
The magnitude of half-filled umklapp scattering, $g_{1/2}$,
   and that of quarter-filled one, $g_{1/4}$, 
   are given by \cite{Tsuchiiz}
\begin{eqnarray}
g_{1/4} &=& \frac{1}{(2\pi\alpha)^2} \, 
\frac{A^4 \, a^5}{v_{\rm F}^2} \, U^2 (U-4V)
,
\label{eq:g1/4}
\end{eqnarray}
   and $g_{1/2} =B  Ua $ in the lowest order where
   $A \equiv [1-(t_d/t)^2]/[1+(t_d/t)^2]$ and
   $B \equiv [2t_d/t]/[1+(t_d/t)^2]$.
We neglect the spin-charge coupling term 
   $U^3 \cos 4\theta_+ \cos 2\phi_+$ due to 
   its large scaling dimension \cite{Yoshioka}.
The order parameters
   of the $4k_{\rm F}$ charge density wave (CDW) state and
   the $4k_{\rm F}$ bond ordered wave (BOW) state
   are given by\cite{Yoshioka}
\begin{eqnarray}
\mathcal{O}_{4k_{\rm F}\mbox{-}\rm CDW}(x)  
&\propto&
(-1)^{x/a}
\cos  2\theta_+(x)
, 
\\
\mathcal{O}_{4k_{\rm F}\mbox{-}\rm BOW} (x)
&\propto&
(-1)^{x/a}
\sin  2\theta_+(x)
, 
\end{eqnarray}
where $k_{\rm F}=\pi/4a$.

The spin degrees of freedom becomes gapless due to the 
   marginally irrelevance of the $g_{s}$-term. 
Therefore we focus hereafter on the charge degrees of freedom only.
In the case of $t_d=0$ (i.e., $g_{1/2}=0$),
   it has been shown by the RG analysis that 
   the system undergoes the
   Kosterlitz-Thouless transition by increasing the 
   nearest-neighbor repulsion \cite{Yoshioka} where
   the gapped phase corresponds to
   the CO insulator.
In the case of $t_d\neq 0$,
the Hamiltonian $H_c$ has two possible ground states with $\langle
   \theta_+\rangle = 0,\pi/2$ ($4k_F$-CDW, dominated by $g_{1/4}<0$)
    or  $\langle \theta_+\rangle = \pi/4$  ($4k_F$-BOW, dominated by
$g_{1/2}$). These two ground states are incompatible, so we expect a
phase transition of the Ising type, as discussed in \cite{Fabrizio}.

By following the method of Ref.~\cite{Fabrizio},
   we analyze the DSG model
   Eq.~(\ref{eq:phaseHamiltonian}).
By using the Jordan-Wigner transformation,
   the vertex operators
   are expressed in terms of the order ($\sigma_1$ and $\sigma_2$) 
   and disorder ($\mu_1$ and $\mu_2$) parameters of the 
   two Ising models as follows \cite{Gogolin_1d_book,Fabrizio}: 
\begin{eqnarray}
\sin 2\theta_+ &=& \sigma_1^z \, \sigma_2^z
,
\label{eq:sin2theta}
\\
\cos 2\theta_+ &=& \mu_1^z \, \mu_2^z
,
\\
\cos 4\theta_+ &=& \frac{1}{2} (\sigma_1^x + \sigma_2^x)
.
\end{eqnarray}
The charge part of the Hamiltonian $H_c$ can be expressed in 
   terms of the quantum Ashkin-Teller model which
   consists of two quantum Ising  spin chains 
   coupled by a four-spin interaction term \cite{Fabrizio}:
\begin{eqnarray}
H_c &=&
H_{\rm QI}[\sigma_1,\sigma_2]
+ H_{\rm AT} [\sigma_1,\sigma_2]
+ H_h [\sigma_1,\sigma_2]
,
\label{eq:DQAT}
\end{eqnarray}
   where 
   $\sigma_l$ are the Pauli matrices of $l$-th chain $(l=1,2)$.
In Eq.~(\ref{eq:DQAT}),
   $H_{\rm QI}$ is given by
\begin{eqnarray}
H_{\rm QI} 
&=&
-\sum_{l=1,2} \sum_n 
\Bigl[
  J \,  \sigma_l^z(n) \, \sigma_l^z(n+1) + \Delta \, \sigma_l^x(n)
\Bigr]
,
\label{eq:Hqi}
\end{eqnarray}
   where $n$ denotes the lattice site on the respective chain.
This Hamiltonian describes two independent
   quantum Ising 
   (QI) chains in a transverse magnetic field ($\Delta$).
In Eq.~(\ref{eq:DQAT}),
  $H_{\rm AT}$ is the Ashkin-Teller interaction term given by
\begin{eqnarray}
H_{\rm AT} 
&=&
J' \sum_n 
\Bigl[
  \sigma_1^z(n) \, \sigma_1^z(n+1) \,
  \sigma_2^z(n) \, \sigma_2^z(n+1) 
\nonumber \\ && \hspace*{1cm} {}
 + \sigma_1^x(n)  \, \sigma_2^x(n)
\Bigr]
,
\end{eqnarray}
   where this term describes  a four-spin interaction.
It is noted that this interaction term takes a {\it self-dual} form,
   thus this term does not make the system non-critical.
Finally,
   $H_h$ is a ``magnetic field''-type coupling, given by
\begin{eqnarray}
H_h 
&=&
-h \sum_n 
  \sigma_1^z(n) \, 
  \sigma_2^z(n) 
.
\end{eqnarray}
By using the Jordan-Wigner transformation,
   the parameters in Eq.~(\ref{eq:phaseHamiltonian}) and
   those in Eq.~(\ref{eq:DQAT}) can be related as 
  $v_\rho = 2J a [ 1- \left(2J'/\pi J \right)^2 ]^{1/2}$,
  $K_\rho = (1/8)
    [ \{1+(2J'/\pi J)\}/\{1-(2J'/\pi J)\} ]^{1/2}$,
  $g_{1/2}= 2a  h$ and
  $g_{1/4} = - 4a (\Delta-J ) $.
Note that the sine-Gordon model, Eq.~(\ref{eq:phaseHamiltonian}), 
   with $K_\rho=1/8$, $g_{1/2}= 0$ and $g_{1/4} \neq 0$, is equivalent 
   to a model of two degenerate massive Majorana fermions,
   i.e., two decoupled non-critical Ising models \cite{Gogolin_1d_book},
   and,  when the nonlinear term vanishes ($g_{1/4}\to0$), 
   two Ising models
   become critical simultaneously, i.e., critical properties
   are described by $c=1$ CFT.

The origin of the Ising criticality ($c=1/2$ CFT) in the case
   of $g_{1/2}\neq 0$ 
   can be captured easily 
   by considering the limit of large $h$ \cite{Fabrizio}.
In this case, the Ising spins are
   locked as $(\sigma_1,\sigma_2)=(1,1)$ or $(-1,-1)$
   where $\sigma_i$ is the eigenvalue of $\sigma^z_i$.
This fact implies that 
   the ``relative'' degrees of freedom 
   becomes locked (i.e., this freedom is always non-critical) 
   while the ``total'' degrees of freedom 
   asymptotically decouples  and 
   the system can become
   critical where the critical properties are described by $c=1/2$ CFT.
In order to analyze the physical quantities on and near 
   the criticality, 
   we change the basis into ``total''-``relative'' representation,
   where the relative Ising mode is described by 
   $\tau=\sigma_1\, \sigma_2$, while the total mode 
   by  e.g., $\sigma=\sigma_1$. 
The operator identities are given by \cite{Fabrizio}
\begin{eqnarray}
\begin{array}{ll}
   \sigma^z = \sigma_1^z , 
&
   \tau^z = \sigma_1^z  \, \sigma_2^z ,
\\
   \sigma^x=\sigma_1^x \, \sigma_2^x , \hspace*{.2cm}
&
   \tau^x=\sigma_2^x .
\end{array}
\end{eqnarray}
In terms of these spin variables, 
    Eq.~(\ref{eq:DQAT}) is rewritten as
   $H_c = H_\sigma + H_\tau + H_{\sigma \tau}$ where
\begin{eqnarray}
H_\sigma
&=& \sum_n
\Bigl[
  -J \, \sigma^z(n) \, \sigma^z(n+1) 
  +J' \, \sigma^x(n)
\Bigr]
,
\\
H_\tau
&=& \sum_n
\Bigl[
   J' \, \tau^z(n) \, \tau^z(n+1) 
  -\Delta \, \tau^x(n)
  -h \, \tau^z(n)
\Bigr]
,
\\
H_{\sigma \tau}
&=& -\sum_n
\Bigl[
   J \, \sigma^z(n) \, \sigma^z(n+1) \, \tau^z(n) \, \tau^z(n+1) 
\nonumber \\ && \hspace*{1.5cm} {}
   + \Delta \, \sigma^x(n) \, \tau^x(n)
\Bigr]
.
\end{eqnarray}
The Ising transition corresponds the situation 
   when the $\sigma$-degrees of freedom go massless
   at $\Delta \approx \sqrt{h(2J+J')}$ \cite{Fabrizio}.
With no regard to the Ising transition in the $\sigma$-chain,
   the $\tau$-chain remains frozen in a 
   configuration with both $\langle \tau^z \rangle$ and 
   $\langle \tau^x \rangle$ 
   nonzero.
Here we introduce the spin variables $\mu$ and $\nu$
  which are dual to
  $\sigma$ and $\tau$, respectively, by 
  the Kramers-Wannier transformation:
  $\mu^z(n+1/2) \equiv \prod_{m=1}^n \sigma^x(m)$,
  $\mu^x(n+1/2) \equiv \sigma^z(n)  \sigma^z(n+1)$,
  $\nu^z(n+1/2) \equiv \prod_{m=1}^n \tau^x(m)$ and
  $\nu^x(n+1/2) \equiv \tau^z(n)  \tau^z(n+1)$.
The spin variables $\mu$ and $\nu$ can be related with
    $\mu_1$ and $\mu_2$ by 
   $\mu^z=\mu_1^z \, \mu_2^z$,
   $\mu^x=\mu_1^x$, 
   $\nu^z=\mu_2^z$ and
   $\nu^x=\mu_1^x \, \mu_2^x$.
The expectation value of $\mu^z$
   becomes finite ($\langle \mu^z\rangle\neq 0$) in the case that
   $\sigma$-chain is in the disordered phase 
   ($\langle \sigma^z\rangle=0$), 
   while becomes zero ($\langle\mu^z\rangle= 0$) in the ordered phase
   ($\langle \sigma^z\rangle \neq 0$).
The order parameter of $4k_{\rm F}$-CDW
   is given in terms of the spin variable by
   $\mathcal{O}_{4k_{\rm F}\mbox{-}\rm CDW} 
    \propto \cos 2\theta_+  =  \mu_1^z \, \mu_2^z=\mu^z$, and then 
   one finds $\langle \mathcal{O}_{4k_{\rm F}\mbox{-}\rm CDW} \rangle 
    \propto \langle \mu^z \rangle \neq 0$  in the disorder phase only. 
Therefore
   the transition from the Mott insulator to the
   CO insulator corresponds to this order-disorder
   Ising quantum phase transition
   of the $\sigma$ Ising chain,
   where the disordered phase of the Ising system
   corresponds to the CO insulating state.
Note that the quantity $\langle \mathcal{O}_{4k_{\rm F}\mbox{-}\rm BOW}\rangle$
   takes a finite value
   and shows no anomaly at the transition point,
   since the order parameter of $4k_{\rm F}$-BOW is given in terms
   of the noncritical Ising field:
   $\mathcal{O}_{4k_{\rm F}\mbox{-}\rm BOW}\propto \sin 2\theta_+
    =\sigma_1^z\, \sigma_2^z = \tau^z$.

Now we examine  the dielectric permittivity and the
   optical conductivity at finite temperature.
In the energy scale higher than the 
   gap in the $\tau$-chain, the perturbative analysis 
   in terms of renormalization group method
   works well \cite{Tsuchiiz}.
The complex conductivity 
   $\sigma(\omega)=\sigma_1(\omega)+i\, \sigma_2(\omega)$
   ($\sigma_1\equiv {\rm Re} \, \sigma$ and 
    $\sigma_2\equiv {\rm Im} \, \sigma$)
   can be calculated from
   $\sigma(\omega)=i[\chi(\omega)-\chi(0)]/\omega$,
   where $\chi$ is the retarded current-current correlation 
   function: $\chi(\omega)=-i\int dx \int_0^\infty dt 
   \langle [j(x,t),j(0,0)]\rangle$ $ e^{i\omega t-\delta t}$.
In terms of the complex conductivity, 
   the dielectric permittivity is given by
   $\epsilon(\omega)=\epsilon_1(\omega)+i \, \epsilon_2(\omega)
    =1+(4\pi i/\omega)\, \sigma(\omega)$,
   where $\epsilon_1\equiv {\rm Re} \, \epsilon$ and
   $\epsilon_2\equiv {\rm Im} \, \epsilon$.
By utilizing the RG method, 
   it can be shown that
   the resistivity $\rho(T)$ and the dielectric permittivity
   $\epsilon_1(T)$ 
   behaves as $\rho(T)=1/\sigma_1(T)\propto T^{16K_\rho-3}$
   and $\epsilon_1(T)\propto T^{2-16K_\rho}$ 
   in the case that the quarter-filled umklapp scattering
   is dominant \cite{Tsuchiiz}.
This analysis, however, only holds in the high temperature region.
In the energy scale lower than the gap in the $\tau$-chain, 
   we analyze these physical quantities in the 
   quantum Ising model scheme.
By using the spin variable $\mu$,
   the current operator are given by
   \cite{Fabrizio}
\begin{eqnarray}
j(x) &=& -\frac{1}{\pi} \, \partial_t \theta_+
  \, \to -C \, \partial_t \, \mu(x)
,
\label{eq:current}
\end{eqnarray}
   where $\mu(x)$ is a continuum variable of $\mu^z(n)$ and
   and $C(\propto \langle \tau^z \rangle)$ is a non-universal
   numerical constant.
Here we have used the fact that the $\tau$-field is
   always non-critical
   and has nonzero expectation value. 
We introduce the  response function of the 
    Ising disorder parameter:
\begin{eqnarray}
D^{(R)}(k,\omega) &=& -i \int dx \, \int_0^\infty dt \,
 e^{-ikx+i\omega t} 
\left\langle
   \left[ \mu(x,t) , \, \mu(0,0)\right]
\right\rangle .
\label{eq:responseIsing}
\end{eqnarray}
By using Eq.~(\ref{eq:current}), 
   the current-current correlation function is given 
   by
   $\chi(\omega)=C^2\omega^2 D^{(R)}(0,\omega)$.
Then the real and imaginary parts of the conductivity $\sigma_1$ and
   $\sigma_2$ are given by
  $\sigma_1(\omega) = -C^2 \, \omega \, {\rm Im} 
   D^{(R)}(0,\omega)$ and
  $\sigma_2(\omega) =  C^2 \, \omega \, {\rm Re} 
   D^{(R)}(0,\omega)$.
On the Ising criticality, it is known that
   the response function 
   is given by
   $D_{\rm crit}(r) \propto 1/r^{1/4}$
   where $r=(x^2+\tau^2)^{1/2}$ 
   with the velocity being set to unity.
This power-law behavior is still valid away from 
   the criticality in the short range region $r\ll 1/\Delta_\sigma$,
   where $\Delta_\sigma$ is the gap in the $\sigma$-part of
   the Ising chain.
By using this asymptotic behavior,
   the retarded response functions is given by \cite{Gogolin_1d_book}
\begin{eqnarray}
D_{\rm crit}^{(R)}(0,\omega) 
 &=& 
-\frac{\sin(2\pi \Delta)}{(2\pi T)^{2-4\Delta}} 
  B^2\left(\Delta-i S, 1-2 \Delta \right)
,
\end{eqnarray}
   where $S=\omega/4\pi T$,
   the conformal dimension is $\Delta(=\bar{\Delta})=1/16$ and
   $B(x,y)$ is the beta function given by
   $B(x,y)=\Gamma(x)\Gamma(y)/\Gamma(x+y)$.
The temperature dependence of the dielectric permittivity
   $\epsilon_1(\omega,T)$ with 
   zero and finite frequency is shown in Fig.~1, where $W$ denotes
   the cutoff of the order of the bandwidth and
   for simplicity we have set $C=1$.
The frequency dependence of $\epsilon_1(\omega,T)$ is also shown
   in the inset.
The dielectric permittivity shows strong temperature dependence. 
In high temperature limit, the dielectric permittivity $\epsilon_1$ 
   becomes unity independent of $\omega$,
   but as decreasing temperature
   the huge enhancement can be obtained and
   $\epsilon_1$ takes a maximum 
   at $T\approx \omega$.
At low temperature ($T  \lsim \omega$), $\epsilon_1$ 
   decreases and changes its sign, and finally saturates to a value
   that depends on $\omega$.
The optical conductivity also increases as decreasing temperature, and
   exhibits a maximum at $T\approx \omega$.   
The limiting behaviors of $\epsilon_1$ and $\sigma_1$ 
   are given by
   $\epsilon_1(\omega,T)= 1+ {\rm const.}/T^{7/4}$ and
   $\sigma_1(\omega,T)\propto \omega^2/T^{11/4}$
   for high temperature ($T \gg \omega$),
   and
   $\epsilon_1(\omega,T)\propto - 1/\omega^{7/4}$ and
   $\sigma_1(\omega,T)\propto 1/\omega^{3/4}$ 
   for low temperature ($T \ll \omega$).

\begin{figure}
\begin{center}
\includegraphics[width=8cm]{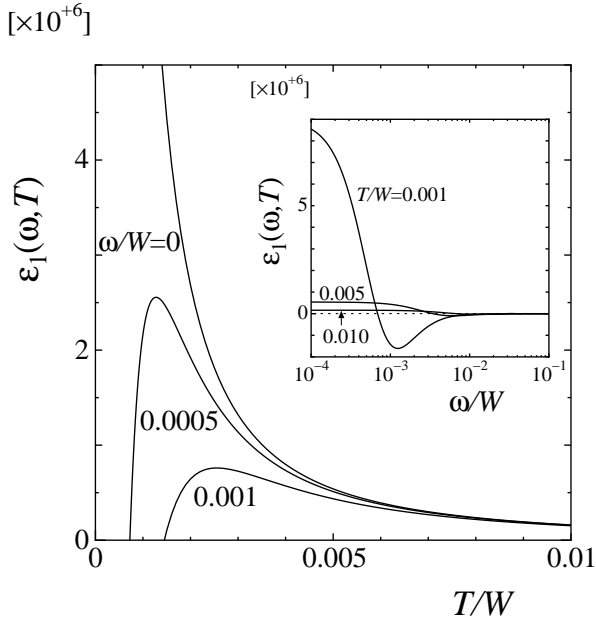}
\end{center}
\caption{ 
The temperature dependence of the dielectric permittivity
   $\varepsilon_1(\omega,T)$ 
   for $\omega/W=0$, 0.0005 and 0.001.
The inset shows the frequency dependence 
   of $\varepsilon_1(\omega,T)$ with fixed temperature
   $T/W=0.001$, 0.005 and 0.01.
}
\end{figure}

Finally we discuss the experimental observation 
   of the (TMTTF)$_2$X salts.
Recent analysis of the dielectric response
   exhibits a strong temperature dependence and a huge enhancement 
   of the dielectric permittivity as decreasing temperature
   \cite{Monceau}.
In our analysis of the temperature dependence of the
    dielectric permittivity,
   we have restricted ourselves to the case on the Ising 
   critical point.
However, the results would be valid 
   in the vicinity of the transition point in the energy region
   above $\Delta_\sigma$.
Therefore the huge enhancement could be interpreted by 
   considering that the system is in the vicinity of the
   transition point from the Mott insulator to the CO insulator.

In summary,  we examined the properties of the
   dimerized extended Hubbard model at quarter-filling
   in one dimension.
It was shown that the phase transition
   from Mott insulating state to the CO insulating state
   in this system
   corresponds to the Ising quantum phase transition described by 
   $c=1/2$ CFT.
We have also show that 
   the dielectric permittivity exhibits a strong enhancement 
   as decreasing temperature with power-law dependence
   at the Ising critical point.

One of the authors (M.T.) thanks A.~Furusaki, T.~Hikihara, H.~Seo,
   Y.~Suzumura and H.~Yoshioka for helpful discussions.


\end{document}